\documentclass{amsart}
\usepackage{amssymb,stmaryrd}
\usepackage{amsfonts}
\usepackage{amstext}

\newcommand{\C}{{\mathbb{C}}}
\newcommand{\R}{{\mathbb{R}}}

\renewcommand{\(}{{(}}

\newcommand{\Tr}{{\rm Trace}}

\newcommand{\cg}{{\mathfrak{g}}}

\newcommand{\Newt}{{\scriptstyle G_N}}

\newcommand{\id}{{\rm id}}

\newcommand{\tens}{\otimes}

\newcommand{\extd}{{\rm d}}

\newcommand{\und}{\underline}

\begin{document}

\title{Cosmological constant from quantum spacetime} 
\keywords{quantum gravity, quantum spacetime, dark energy, noncommutative geometry, differential algebra, quantization conditions,  cosmological constant}

\author{Shahn Majid \& Wen-Qing Tao}
\address{Queen Mary University of London,\\
School of Mathematical Sciences, Mile End Rd, London E1 4NS, UK
}
\email{s.majid@qmul.ac.uk}
\thanks{The first author was on leave at the Mathematical Institute, Oxford, during 2014. The second author was supported by the China Scholarship Council}

\begin{abstract} We show that a hypothesis that spacetime is quantum with coordinate algebra $[x^i,t]=\lambda_P x^i$, and spherical symmetry under rotations of the $x^i$, essentially requires in the classical limit that the spacetime metric is the Bertotti-Robinson metric, i.e. a solution of Einstein's equations with cosmological constant and a non-null electromagnetic field. Our arguments do not give the value of the cosmological constant or the Maxwell field strength but they cannot both be zero. We also describe the quantum geometry and the full moduli space of metrics that can emerge as classical limits from this algebra. \end{abstract}
\maketitle 

\section{Introduction}\label{Sec1}

Recently in \cite{BegMa6} was uncovered a new phenomenon whereby the constraints of noncommutative algebra force the form of quantum metric and hence of its classical limit. Put another way, if a spacetime is quantised, as is by now widely accepted as a plausible model of quantum gravity effects, then this would be visible classically as quantisability conditions\cite{BegMa7} on the classical spacetime metric so as to extend to the quantum algebra. Thus the quantum spacetime hypothesis potentially has strong and observable consequences for GR.

Specifically, \cite{BegMa6} looked 
at the most popular quantum spacetime algebra, the bicrossproduct or Majid-Ruegg model\cite{MaRue} with generators $x^i,t$, $i=1,\cdots , n-1$  and relations
\begin{equation}\label{bicross} [x^i,x^j]=0,\quad [x^i,t]=\lambda x^i\end{equation}
where $\lambda=\imath\lambda_P$ and $\lambda_P$ is a real quantisation parameter, usually assumed in this context to be the Planck time. Here $n=4$ but we will consider other dimensions also. The paper \cite{BegMa6} showed that
in the 2D case the quantisability constraints force a strong gravitational source or an expanding universe depending on a sign degree of freedom. This provided a toy model but
 in  4D the constraints were so strong that there was no fully invertible quantum metric at all. The analysis depended on the differential structure on the algebra and we used the standard one as in \cite{Oec,AmeMa}. 
 
In the present paper we will now consider the same phenomenon  for another natural choice of differential structure on (\ref{bicross}), which we call the `$\alpha$ family' and which  we show, in Section~\ref{Sec2}, is the only good alternative that treats the $x^i$ equally in the sense of rotationally invariant and works in all dimensions. The relations for this differential calculus first appeared in \cite{MKS} as did those for another, which we call the` $\beta$ family' and which generalises the standard one. In our case we come to these same differential calculi out of a systematic classification theory\cite{MaTao:pre} based on pre-Lie algebras.  Remarkably we  then find for the $\alpha$ family, in Section~\ref{Sec3}, that this time there {\em is} a moduli of quantum metrics and in Section~\ref{Sec4} we consider their classical limits and show that in the spherically symmetric case they are all locally of the form $S^{n-2}\times dS_2$ or $S^{n-2}\times AdS_2$ depending on the sign of one of the two curvature-scale parameters $\delta,\bar\delta$. This means that they are the Levi-Bertotti-Robinson metric \cite{Levi,Ber,Rob,Gur}, which has been of interest in a number of contexts in GR and is known to solve Einstein's equation with cosmological constant and Maxwell field. We can write the value of the cosmological constant here as
\[ \Lambda={(n-2)(n-3)\over 2}\delta-q^2 \Newt,\quad q^2\Newt= {1\over 2}\left((n-3)\delta-\bar\delta\right)\]
where $q$ is the Maxwell field coupling in suitable units. In our context $\delta>0$ so that for small $q$ we are forced to $\Lambda>0$.   Moreover, the arguments that force us to this form of metric depend on the structure of the differential algebra when spacetime is noncommutative, which is believed to be a quantum gravity effect. In 2D there is no $S^{n-2}$ factor and being the limit of a quantum metric in the $\alpha$ family in 2D  forces the metric to be de Sitter or anti-de Sitter for some scale $\bar\delta$.

The further noncommutative Riemannian geometry for our quantum metrics in the $\alpha$ family is obtained by the same methods as in \cite{BegMa6} and a brief outline of this is included in the final Section~\ref{SecA} for completeness. We work in this paper with one particular algebra (\ref{bicross}) assumed to be some local model of quantum spacetime. The general analysis at lowest order in $\lambda$, i.e. at the level of a general Poisson structure on spacetime and the constraints on the classical metric metric arising from being quantisable along with it, can be found in \cite{BegMa7}. 

An earlier model where vacuum energy was speculated to arise from noncommutative geometry of the quantum spacetime (\ref{bicross}) was the non-relativistic gravity model in \cite{Ma:new}. A cosmological constant is also needed for quantum Born reciprocity in 3D quantum gravity\cite{MaSch}, which paper also shows how the 3D version of (\ref{bicross}) can arise there.

 \section{Choice of differential structure} \label{Sec2}
 
Differential structure on an algebra means for us a specification of the exterior algebra of `differential forms' or in practice the commutation relations between differentials $\extd x^i,\extd t$ and quantum spacetime coordinates.  The exterior derivative $\extd$ on arbitrary noncommutative functions in the coordinates is then defined by the Leibniz rule. We look for differential structures that are (i) connected, meaning only constant functions are killed by $\extd$ and (ii) translation-invariant with respect to the additive coproduct on (\ref{bicross}). The latter says that as a differential space this is much like $\R^n$ in the same way that a classical manifold has local coordinates where the differentials $\extd x^i,\extd t$ are related to the standard translation-invariant Lebesgue measure. 

Our starting point is a recent theorem \cite{MaTao:pre} that connected translation invariant differential structures of the correct classical dimension on the enveloping algebra of a Lie algebra $\cg$ are in 1-1 correspondence with pre-Lie algebra structures on $\cg$. This means a map $\circ:\cg\tens\cg\to \cg$ such that $x\circ y-y\circ x$ recovers the given Lie algebra bracket and 
\[ (x\circ y)\circ z-(y\circ x)\circ z=x\circ(y\circ z)-y\circ(x\circ z)\]
Moreover, for a Lie algebra like (\ref{bicross}) there is an algebraic method which provides all inequivalent pre-Lie algebra structures, which in the 2D case\cite{Bu} over $\C$ gives 2 distinct families and 3 discrete choices at the algebraic level. The corresponding differential structures in the 2D case are computed in \cite{MaTao:pre} and come out  as
\begin{eqnarray*}&(i):&\   [t,\extd x]=-\lambda \extd x,\quad [t,\extd t]=\lambda\alpha\extd t\\
&(ii):&\  [x,\extd t]=\lambda\beta \extd x,\  [t,\extd x]=\lambda(\beta-1)\extd x,\  [t,\extd t]=\lambda\beta\extd t\\
&(iii):&\   [t,\extd x]=-\lambda \extd x,\quad [t,\extd t]=\lambda(\extd x- \extd t)\\
&(iv):&\  [x,\extd x] =\lambda\extd t,\quad [t,\extd x]=-\lambda \extd x,\quad [t,\extd t]=-2\lambda\extd t\\
&(v):&\  [x,\extd t]=\lambda\extd x, \quad  [t,\extd t]=\lambda(\extd x+ \extd t).\end{eqnarray*}
In each case we have listed only the non-zero commutation relations. Of these, clearly, only (i) and (ii) immediately generalise to all dimensions, namely as the `$\alpha$ family'
\begin{equation}\label{alpha}  [t,\extd x^i]=-\lambda \extd x^i,\quad [t,\extd t]=\lambda\alpha\extd t \end{equation}
and the `$\beta$ family'
\begin{equation}\label{beta} [x^i,\extd t]=\lambda\beta \extd x^i,\quad [t,\extd x^i]=\lambda(\beta-1)\extd x^i,\quad [t,\extd t]=\lambda\beta\extd t  \end{equation}
for the non-zero relations, cf \cite{MKS} where there are some similar relations to these two families. The case $\beta=1$ of the second family is the standard calculus used in \cite{Oec,AmeMa,BegMa6}. It should also be noted that case (v) is equivalent to the standard calculus in 2D in case (ii)  by a change of variables if we allow a sufficient class of functions. Likewise  case (iii) is
equivalent to $\alpha=-1$ in case (i) if we allow a sufficient class of functions. 

We consider only these two families (\ref{alpha}), (\ref{beta}) in what follows: by our above results, they are the only connected translation invariant differential structures in the quantum spacetime (\ref{bicross}) that work in
all dimensions including 2D.  To fully classify all 4D calculi is also possible using the algebraic method for pre-Lie algebras\cite{Bai} and could include more exotic possibilities
but they are unlikely to treat the different $x^i$ equally in the sense of spatial rotations as otherwise they would specialise to 2D.  

\section{Quantum metrics for the $\alpha$ and $\beta$ calculi}\label{Sec3}

In both cases in we will tend to focus on the radial-time sector of the algebra. Here $r=\sqrt{\sum_i x^i{}^2}$ and the inherited relations are $[r,t]=\lambda r$ as well as relations for $\extd r$ of the same form as for $\extd x^i$. In both cases we let
\[ \omega^i=\extd x^i- {x^i\over r}\extd r\]
which commutes with spatial variables. Here the sphere direction variables $z^i={x^i\over r}$ commute with $x^i,t$ according to the relations (\ref{bicross}) and obey $\sum_i z^i{}^2=1$, and $\extd z^i=r^{-1}\omega^i$. The element $r^{-2}\sum_i \omega^i\tens\omega^i$ 
makes sense generally and in our case, for the $\alpha$ and $\beta$ calculi where spatial differentials and functions commute, behaves classically so long as $t,\extd t$ are not involved. Within this sector (or of course in the classical limit $\lambda_P\to 0$) we can use standard polar coordinates, when it becomes
\[  r^{-2}\sum_i \omega^i\tens\omega^i=\extd\Omega^2=\extd\theta\tens\extd\theta+\sin^2(\theta)\extd\phi\tens\extd\phi\]
in 4D. In $n$ dimensions, $\extd\Omega^2$ here is the metric on the unit sphere $S^{n-2}$. We use a more explicit notation for tensor products (over the coordinate algebra) than is usual in GR. Also, we will extend our noncommutative algebra to include say $r^{\pm\alpha},r^{\pm\beta}$ and in the classical limit all smooth functions. We are working as in quantum mechanics with hermitian $x^i{}^*=x^i$ and $t^*=t$, and $r^*=r$. In the classical limit the $*$ becomes complex conjugation of functions and these requirements become that our coordinates are real. The $*$ extends to the differentials with $\extd *=*\extd$. 

For a quantum metric we take something of the form $g=g_{\mu\nu}\extd x^\mu\tens_1\extd x^\nu$ where $x^0=t$ and $x^i$ are the spatial coordinates, the coefficients $g_{\mu\nu}$ are elements of the quantum coordinate algebra and the subscript 1 reminds is that this is the quantum tensor product, i.e., over the quantum coordinate algebra. We require the quantum metric to be invertible in the sense of an inner product $(\ ,\ )$ on the space of 1-forms, which behaves well (is` strongly tensorial') with respect to multiplication by co-ordinates in the sense\cite{BegMa6},
\[  f(\omega ,\eta )=(f\omega ,\eta ),\quad (\omega f,\eta )=(\omega ,f\eta ),\quad(\omega ,\eta f)=(\omega ,\eta )f\] for all elements $f$ of the quantum coordinate algebra and all 1-forms $\omega,\eta$. It is shown in \cite{BegMa6} that this requires $g$ to commute with elements of the quantum coordinate algebra. We also require that
$g$ is `quantum symmetric' in the sense $\wedge(g)=0$. The quantum wedge product here is an extension of the 1-forms to an associative product on forms of all degree and to which $\extd$ extends. Here the basic one forms $\extd x^i,\extd t$ obey the usual exterior or Grassmann algebra (they anticommute). Finally, we need a condition that expresses reality of the metric coefficients, which we express as \cite{BegMa6,BegMa5}
\begin{equation}\label{reality} (*\tens_1 *){\rm flip}(g)=g\end{equation}
where `flip' swaps the factors of $\tens_1$.  We will in practice omit the subscript on the tensor product as this should be clear from context.

\subsection{Quantum metrics for the $\alpha$-calculus}\label{Sec3.1}

For the $\alpha$ family (\ref{alpha}) we consider a quantum metric of the arbitrary form
\[g=\sum^{n-1}_{i,j}a_{ij}\extd x^i\tens \extd x^j+\sum_i^{n-1} b_i (\extd x^i\tens \extd t+\extd t\tens\extd x^i)+c\extd t\tens\extd t,\] where the coefficients $a_{ij},b_i,c$ are all elements in the quantum spacetime algebra and obey $a_{ij}=a_{ji}.$ This form is dictated by `quantum symmetry' in the form $\wedge (g)=0.$ Using the Leibniz rule and the relation (\ref{alpha}) we have
\begin{eqnarray*}
{} [g,t]&=&\sum_{i,j}^{n-1} ([a_{ij},t]+2\lambda a_{ij})\extd x^i\tens\extd x^j \\
&&+ \sum_i^{n-1} ([b_i,t]-\lambda(\alpha-1)b_i)(\extd x^i\tens \extd t+\extd t\tens\extd x^i)\\
&&+ ([c,t]-2\lambda\alpha c)\extd t\tens \extd t.\\
{} [g,x^k]&=&\sum_{i,j}^{n-1} [a_{ij},x^k]\extd x^i\tens\extd x^j+\sum_i^{n-1}[b_i,x^k](\extd x^i\tens \extd t\\
&&+\extd t\tens\extd x^i)+[c,x^k]\extd t\tens \extd t.
\end{eqnarray*}
This means that $g$ central amounts to
\[
[a_{ij},t]=-2\lambda a_{ij},\ \forall i,j,\quad [b_i,t]=\lambda(\alpha-1) b_i,\ \forall i,\]
\[ [c,t]=2\lambda\alpha c,\quad [a_{ij},x^k]=0,\ \forall i,j,k,\]
\[ [b_i,x^k]=0,\ \forall i,k,\quad [c,x^k]=0,\ \forall k.
\]
By solving this, we see that our requirements are that $a_{ij}, b_i, c$ are all functions only of $x$ and have scaling degree $-2,\alpha-1,2\alpha$ respectively.  Hence there is a larger moduli of metrics for this differential calculus; we just have to make sure that the coefficients are homogeneous of the appropriate degree. 

If we look among spherically symmetric quantum metrics, which seems natural from the form of the algebra (\ref{bicross}), then  we have
\begin{eqnarray}\label{quantalpha} g&=& \delta^{-1} r^{-2}\sum_i\omega^i\tens\omega^i+ a r^{-2}\extd r\tens\extd r\nonumber\\
&& + br^{\alpha-1}(\extd r\tens\extd t+\extd t\tens\extd r)+c r^{2\alpha} \extd t\tens\extd t\end{eqnarray}
for $\delta, a,b,c\in\R$, which by the above is central. Here $\delta>0$ could be normalised to $\delta=1$ but we have refrained from this as it is dimensionful with dimensions of inverse square length. The quantum metric is  quantum symmetric and obeys the `reality' condition (\ref{reality}) given that $r$ commutes with $\extd x^i,\extd t$ in this calculus.  

\subsection{Quantum metrics for the $\beta$-calculus}\label{Sec3.2}

The $\beta$ family (\ref{beta}) contains the standard calculus at $\beta=1$ and we find basically the same result as for that in \cite{BegMa6}. We will omit the details and the proof as the methods are the same but the result is:  {\em For the $\beta$ family calculi in dimension $n>2$ there are no central quantum metrics among a reasonable class of coefficient functions.}

One can, however, consider metrics that  are spherically symmetric and commute with functions of $r,t$. To do this let us first note that the elements
\[u=r^{\beta-1}\extd r,\quad v=r^{\beta-1}(r\extd t-\beta t\extd r),\quad r^{\beta-1}\omega_i\] 
commute with $r,t$.  Also 
\[ u^*=u,\quad v^*=\lambda \beta(\beta-2)u+v,\quad \omega_i^*=\omega_i\]
using the commutation relations.  Looking in the 2D $r-t$ sector, the element 
\[
g_{2D}=v^*\tens v+\beta\lambda (u\tens v-v^*\tens u)-\gamma_1(u\tens v+v^*\tens u)+\gamma_2 u\tens u
\]
then manifestly commutes with $t,r$ and is `real' in the hermitian sense provided $\gamma_1,\gamma_2$ are real, and also manifestly obeys $\wedge(g)=0$. 
Now let $t'=t+\frac{\gamma_1}{\beta},$ so $\extd t'=\extd t, v'=r^{\beta-1}(r\extd t'-\beta t'\extd r)=v-\gamma_1 u,$ thus
\[
g_{2D}=v'^*\tens v'+\beta\lambda(u\tens v'-v'^*\tens u)+\gamma u\tens u,
\]
where $\gamma=\gamma_2-{\gamma_1}^2$ is a real parameter. Therefore we can assume that the time variable has been shifted to eliminate the $\gamma_1$ term as the expense of the $\gamma_2$ term. We now combine this information with the angular part of the metric, so 
\begin{equation}\label{quantbeta} g= r^{2\beta-2}\sum_i \omega^i\tens\omega^i +a u\tens u+b v^*\tens v+\beta b \lambda(u\tens v-v^*\tens u),\end{equation}
for $a,b\in\R$, $a,b\ne 0$, 
commutes with $r,t$. One could insert an overall normalisation to fix the dimensions of $g$. The additional angular term commutes, has zero wedge product and obeys the reality condition, so these features all still hold for $g$.  This metric generalises the one in \cite{BegMa6} from the case $\beta=1$. Using the same methods as in \cite{BegMa6} we can show that up to a shift in the $t$ variable, this is the most general form of spherically symmetric metric  that commutes with $r,t$ and involves a reasonable class of functions.

\section{Classical limits}\label{Sec4}

We now look at the classical limits of the spherically symmetric quantum metrics allowed in Section~\ref{Sec3}.

\subsection{Emergence of Bertotti-Robinson metric from the $\alpha$ family}\label{Sec4.1}

Here we look at the classical limit of the metric in Section~\ref{Sec3.1}, namely
\begin{eqnarray}\label{classalpha} g&=&\delta^{-1}\extd\Omega^2+ a r^{-2}\extd r\tens\extd r\nonumber\\
&&+ br^{\alpha-1}(\extd r\tens\extd t+\extd t\tens\extd r)+c r^{2\alpha} \extd t\tens\extd t\end{eqnarray}
where $a,b,c\in\R$, $\delta>0$ and we need $b^2-ac>0$ for a Minkowski signature. 
The first thing we do is define the combination 
\[ \overline{\delta}={c\alpha^2\over b^2-ac}\]
and compute for $n\ge 3$ that the Einstein tensor  is 
\begin{equation}\label{Gmod} G=-{(n-2)(n-3)\over 2}\delta g +((n-3)\delta-\overline{\delta})\delta^{-1}\extd\Omega^2.\end{equation}
We also mention the scaler curvature 
 \begin{equation}\label{Smod} S=(n-2)(n-3)\delta+ 2\overline{\delta}\end{equation}
 and that the metric is conformally flat for $n<4$, while for $n=4$ it is conformally flat when $\delta+\bar\delta=0$.

Our first observation is that this $G$ can never match a perfect fluid other than the vacuum energy case given by $(n-3)\delta=\bar\delta$. This is because  the one-upper index Einstein tensor $\und G$ is diagonal in our coordinate basis with eigenvalues 
\begin{equation}\label{Gevals}-{(n-2)(n-3)\over 2}\delta,\quad -\overline{\delta}-{(n-3)(n-4)\over 2}\delta\end{equation}
where the first eigenspace is spanned by the $t,r$ directions and the other eigenspace is spanned by  the angular directions. Now if the two eigenvalues of $\und G$ are distinct then we cannot have $\und G=8\pi\Newt(p\, \id+(p+\rho)U\tens u)$  for a timelike 1-form $u$ and associated vector field $U$ because this would require $u$ to have only one non-zero entry (since otherwise $U\tens u$ would have off-diagonals) and in that case adding $U\tens u$ can only change the eigenvalue in a 1-dimensional subspace, contradicting the equality of the eigenvalues in the $r,t$ subspace. 

Next we define a Maxwell field strength 
\begin{equation}\label{Max} F= q\sqrt{b^2-ac}\, r^{\alpha-1}(\extd t\tens\extd r-\extd r\tens\extd t)\end{equation}
when viewed as a tensor product of 1-forms. Its stress energy tensor 
\[ T_{\mu\nu}={1\over 4\pi}(F_{\mu\alpha}F^\alpha{}_\nu-{F^2\over 4}g_{\mu\nu})\]
works out as 
\[T=-{q^2\over 4\pi }({g\over 2}-\delta^{-1}\extd\Omega^2)\]
after a short computation. Here $F^2=-2q^2$ so that when present (i.e. when $q\ne 0$) the electromagnetic type is non-null. Comparing with (\ref{Gmod}),  we obey Einstein's equation with cosmological constant $\Lambda$ if we set
\[ \Lambda={(n-2)(n-3)\over 2}\delta-q^2\Newt,\quad q^2 \Newt ={1\over 2}\left((n-3)\delta-\bar\delta\right)\]
which entails $\bar\delta\le (n-3)\delta$, with the case of equality being the vacuum energy solution already noted. This also
implies that 
\begin{equation} \Lambda={1\over 2}\left((n-3)^2\delta+\bar\delta\right)\end{equation}
\begin{equation} ({n-2\over 2})\bar\delta\le \Lambda\le {(n-2)(n-3)\over 2}\delta.\end{equation}
 In the important case of $n=4$ we see that the cosomological constant vanishes exactly in the conformally flat case $\bar\delta=-\delta$ while the Maxwell field strength vanishes exactly in the case $\bar\delta=\delta$. 

These computations are a generalisation of \cite{Gur} for the standard form of Berttoti-Robinson metric if we take 
$\alpha=-1, \delta=q=q'{}^{-2}, a=c_0^2 q'{}^2, b=0, c=-q'{}^2,\bar\delta=-c_0^{-2} q'{}^{-2}$ in terms of the notation there,  denoting $q$ in \cite{Gur} as $q'$ to avoid confusion with our $q$.
On the other hand we now show that all cases of (\ref{classalpha}), even as we vary $\alpha$, are locally equivalent 
to the Bertotti-Robinson metric up to a change of variables, i.e the moduli space is in fact only the two real parameters
$\delta,\bar\delta$. We treat the different signs of $\bar\delta$ separately. 

(i) If $\bar\delta>0$ then this implies $c,a+{\alpha^2\over\overline{\delta}}>0$. We define  a change of variables
\[ t'={\alpha\over\sqrt{\overline{\delta}}}\ln r,\quad r'=\sqrt{c}t- {\sqrt{a+{\alpha^2\over\overline{\delta}}}\over \alpha r^\alpha}\]
 when $b>0$ and the opposite sign in the 2nd term of $r'$ when $b<0$. Then our metric becomes
\begin{equation}\label{ds} g=\delta^{-1} \extd\Omega^2+e^{2 t'\sqrt{\overline{\delta}}}\extd r'^2-\extd t'^2.\end{equation}
which is a known form of the Bertotti-Robinson metric. Indeed,  comparing to 2D de Sitter in the flat slicing
\[ g_{dS}=e^{2t\sqrt{\delta}}\extd x-\extd t^2\]
we see that the metric is that of a part of $S^{n-2}\times dS_2$ with respective curvature scales $\delta, \bar\delta$.

(ii) If $\overline{\delta}=0$ we have $c=0$ and we use a different change of variables: if $a>0$ say,
\[ r'=\alpha r^\alpha-{1\over\alpha r^\alpha}+{2 b\over a}  t,\quad t'=\alpha r^\alpha+{1\over\alpha r^\alpha}-{2 b\over a} t,\]
\[g=\delta^{-1}\extd\Omega^2+{a\over4 \alpha^2}\left(\extd r'{}^2-\extd t'{}^2\right).\]
If $a<0$ we use the same but swap the roles of $t',r'$. We have the metric of a part of $S^{n-2}\times\R^2$ with sphere curvature scale $\delta$.

(iii) If $\overline{\delta}<0$ then $c<0, a+{\alpha^2\over\overline{\delta}}<0$ and we define
\[  r'={\alpha\over\sqrt{-\overline{\delta}}}\ln r,\quad t'=\sqrt{-c}t+{\sqrt{-a-{\alpha^2\over\overline{\delta}}}\over \alpha r^\alpha}\]
 when $b>0$ and the opposite sign in the 2nd term of $t'$ when $b<0$. Then our metric becomes
\begin{equation}\label{grav} g=\delta^{-1} \extd\Omega^2-e^{2 r'\sqrt{-\overline{\delta}}}\extd t'^2+\extd r'^2. \end{equation}
This should be compared with 2D anti-de Sitter space metric, a part of which in certain coordinates can be written as
\[ g_{AdS}=-e^{2 v\sqrt{-\overline{\delta}}}\extd t^2 +\extd v^2.\]
We see that the metric is that of a part of $S^{n-2}\times AdS_2$ with respective curvature scales $\delta, \bar\delta$.

{\em In summary}, up to local changes of coordinates, classical metrics which are classical limits of quantum metrics for the $\alpha$ calculus 
and which are spherically symmetric, are given by two parameters $\delta,\bar\delta$ and equivalent to  the Bertotti-Robinson metric.

If we drop the spherical symmetry assumption, i.e. we just ask for classical metrics that are limits of quantum ones, then we have the allowed form
\begin{eqnarray*} g&=&h+r^{-1}(\eta\tens\extd r+\extd r\tens\eta)+r^\alpha(\zeta\tens\extd t+\extd t\tens\zeta) \\
&&+a r^{-2}\extd r\tens\extd r\nonumber+ br^{\alpha-1}(\extd r\tens\extd t+\extd t\tens\extd r)+c r^{2\alpha} \extd t\tens\extd t\end{eqnarray*}
due to the degree requirements in Section~\ref{Sec3.1} and our polar decomposition, where $h=h_{ij}(z)\extd z^i\tens\extd z^j$ is now a general metric on $S^{n-2}$,  $a,b,c$ are now functions in $S^{n-2}$, and $\eta,\zeta$ are further possible 1-forms on $S^{n-2}$. The Einstein tensor is now typically much more complicated. This polar decomposition also applies in the quantum case of Section~\ref{Sec3.1}.

We note in passing that in $n=4$ we can consider classical metrics like (\ref{classalpha}) but replace $S^2$ with $\delta^{-1}\extd \Omega$ by   a general surface $\Sigma$ with metric $h_\Sigma$, and set $\delta=S_\Sigma/2$ according to its Ricci scaler curvature. We keep $a,b,c$ constant. Then the calculations above go through in the same way and the Einstein tensor suggestively matches  the stress energy of a Maxwell field  (\ref{Max}) and $\Lambda,q^2\Newt$ given by the same formulae as before but typically now varying on $\Sigma$ on account of $\delta$ varying. The constant case $H^2\times dS_2$ or $H^2\times AdS_2$ where we use the hyperboloid $H^2$ with curvature scale $\delta<0$ gives constants and completes the Bertotti-Robinson family.

\subsection{Emergence of flat metric from the $\beta$ family}\label{Sec4.2}

The results are again much the same as in \cite{BegMa6}. The conceptually new result is that for a different choice of $\beta$ we can, however, see flat spacetime as emerging from our algebraic considerations. 

We will again be interested in matching to a perfect fluid. If so then this implies
\[ \und G.U=-8\pi \Newt \rho U\]
so that $U$ is a timelike eigenvector with eigenvalue $-8\pi \Newt \rho$. We look for this first as a necessary but not sufficient condition for matching to a perfect fluid. Having identified the possible
values of $-8\pi \Newt \rho$ we look in its eigenspace for a timelike vector $U$ such that the original Einstein equation holds. In this case
\[ p={1\over 3}\left(\rho+{\Tr\, \und G\over 8\pi \Newt }\right)\]
is also necessary and we see for what parameter values $G$ now obeys Einstein's equation. We focus on the 4D case. The 2D case of course automatically has $G=0$. 

The classical limit of the metric in Section~\ref{Sec3.2} is
\[ g= r^{2\beta-2}( r^2\extd\Omega^2+ (a+\beta^2 b t^2)\extd r^2-2 b r t\extd t\extd r+b r^2 \extd t^2). \]
From the determinant of $g$, we need $a +( \beta^2-1)b t^2$ and $b$ to have opposite sign to have the possibility of Minkowski signature which, looking at small $t$ means $a,b$ have opposite sign and looking for large $t$ means $\beta^2\le 1$. We find Ricci scaler
\[ S=2{ a (a - \beta(4\beta+3) )+ ( \beta^2-1)b t^2 (2a-3\beta^2 + (\beta^2-1)b t^2) 
  \over  r^{2 \beta} (a + (\beta^2-1)b t^2)^2}\]

As far as matching a perfect fluid is concerned, there are three distinct eigenvalues of the Einstein tensor. One of these gives us
\[ 8\pi \Newt  \rho=-{\beta(\beta+1)(2a +\beta(\beta-1) b t^2)\over r^{2\beta}(a+( \beta^2-1) b t^2)^2}\]
with null space of $\und G+8\pi \Newt  \rho$ being the angular directions. These aren't timelike. The other choices of eigenvalue have
\begin{eqnarray*} 8\pi \Newt \rho&=& \frac{a^2-\beta(2\beta+1)a+(\beta^2-1)b t^2 ((\beta^2-1)b t^2-2\beta^2+2a)}{r^{2\beta}(a+ (\beta^2-1)b t^2)^2} \\
&& \pm\frac{\beta(\beta+1)\sqrt{ a^2+6 \beta(\beta-1) a b  t^2+ (\beta-1)^2 \beta (4+5 \beta)b^2 t^4}}{r^{2\beta}(a+ (\beta^2-1)b t^2)^2}
\end{eqnarray*}
and of these only the + sign has a timelike vector in the null space of $\und G+8\pi \Newt \rho$.  Taking this, we then require Einstein's equation to hold and this fixes $a=\beta^2,-\beta(\beta+2)$ (looking at the $t=0$ term in an expansion of Einstein's equation) and this then fixes  $\beta=\pm 1$ (looking at higher powers of $t$). 

The case $\beta=1$ is the case covered in \cite{BegMa6} and these are the two choices $a=1 (b<0)$ 
and $a=-3 (b>0)$ in this case that were found there, respectively a strong gravitational source with positive pressure but zero density and a cosmological solution with negative pressure and positive density (quintessence ratio $-{1\over 2}$). 

The case $\beta=-1$ has allowed values $a=1$ and this turns out to correspond precisely to flat space. This is realised by the change of variables 
\[ r'=r^{-1},\quad x'{}^i=r^{-2}x^i,\quad  t'=r^{-1}t\]
given in \cite{BegMa6} which
rendered $g$ manifestly flat up to a a factor $r'{}^{-4}=r^4$. The above metric is $r^{-4}$ times the metric in \cite{BegMa6} so now becomes the flat metric $g=\extd  x'{}^i{}^2+b \extd t'{}^2$ where $b<0$. 

We conclude that in the 3-parameter space of $a,b,\beta$ there are precisely three cases where the Einstein tensor matches a perfect fluid, namely the two cases already in \cite{BegMa6} where $\beta=1$ and
the new case $\beta=-1$ which has the flat metric when $a=1$. Thus, while the flat metric does not extend to a full quantum metric, it does extend to the class that partially commutes, namely with $r,t$.

\section{Quantum geometry}\label{SecA}

Finally, we show that the quantum metric found in Section~\ref{Sec3.1} indeed leads to quantum Riemannian geometry in the formalism of \cite{BegMa5,BegMa6} and we find, remarkably, that the change of variables that diagonalised our classical metric in Section~\ref{Sec4.1} also provides canonically conjugate variables for the quantum algebra, i.e. the quantum spacetime in the radial-time sector is a Heisenberg algebra as in ordinary quantum mechanics.   

We start with the $n=2$ case so we are doing `quantum de Sitter space', leaving out the $\extd\Omega^2=r^{-2}\sum\omega^i\tens\omega^i$ term from the quantum metric (\ref{quantalpha}). In the classical limit in  Section~\ref{Sec4.1} we used a change of variables (\ref{ds}) to convert this to de Sitter spacetime for some scale $\overline{\delta}=c\alpha^2/(b^2-ac)$. We focus on $\overline{\delta}>0$ but the anti-de Sitter case can be handled similarly.  

Since $r$ commutes with both $\extd t$ and $\extd r$ the change of variable we used classically works just as well in the quantum case. So working in the quantum algebra as in Section~\ref{Sec3.1} we set
\[ T={\alpha\over\sqrt{\overline{\delta}}}\ln r,\quad R=\sqrt{c}t-{\sqrt{a+{\alpha^2\over \overline{\delta}}}\over \alpha r^\alpha},\]
\[ g=e^{2T\sqrt{\overline{\delta}}}\extd R\tens\extd R-\extd T\tens\extd T\]
where 
\[ \extd T={\alpha\over r\sqrt{\overline{\delta}}}\extd r,\quad \extd R=\sqrt{c}\extd t+{ \sqrt{a+{\alpha^2\over\overline{\delta}}}\over r^{\alpha+1}}\extd r\]
works out in just the same way. But note that $[f(r),t]=\lambda r f'(r)$ in view of the commutation relations $[r,t]=\lambda r$. Hence in terms of the new variables we have
\[ [T,R]=[{\alpha\over\sqrt{\overline{\delta}}}\ln r,\sqrt{c}t]=\lambda',\quad \lambda'=\lambda\sqrt{b^2-ac}.\]
In other words, $T,R$ are a canonical conjugate pair with Heisenberg relations between them, for a modified parameter $\lambda'$. Similarly, using the relations of the $\alpha$ family calculus we find
\begin{eqnarray*} {}[R,\extd T]&=&[\sqrt{c}t,{\alpha\over r\sqrt{\overline{\delta}}}\extd r]={\alpha\sqrt{c}\over\sqrt{\overline{\delta}}}[t,{\extd r\over r}]=0\\
{} [R,\extd R]&=&[\sqrt{c}t,\sqrt{c}\extd t+{ \sqrt{a+{\alpha^2\over\overline{\delta}}}\over r^{\alpha+1}}\extd r]\\ 
&=&\lambda c\alpha\extd t+\lambda \sqrt{a+{\alpha^2\over\overline{\delta}}}{\alpha\over r^{\alpha+1}}\extd r=\lambda'\sqrt{\overline{\delta}}\extd R\end{eqnarray*}
and more obviously $ [T,\extd T]=0$,  $[T,\extd R]=0$. So we have a closed algebra of the $R,T$ and their differentials which we now adopt (we can regard the passage between the two sets of variables as formal). This is a nonstandard differential calculus on the familiar Heisenberg algebra.  We also have $R^*=R$ and $T^*=T$ as our change of variables involved only real coefficients  and we suppose that we can extend our Heisenberg algebra to include exponentials of $T$, for example in some operator realisation. We can check our calculations by seeing that $g$ is indeed central:
\begin{eqnarray*}{} [ R,g]&=&[R,e^{2T\sqrt{\overline{\delta}}}]\extd R\tens\extd R+e^{2T\sqrt{\overline{\delta}}}([R,\extd R]\tens\extd R+\extd R\tens[R,\extd R])=0\end{eqnarray*}
using the Heisenberg relations for the first term and the $[R,\extd R]$ relations for the 2nd term. 

Next we write down
the quantum Levi-Civita connection
\[ \nabla \extd R=-\sqrt{\overline{\delta}}(\extd R\tens\extd T+\extd T\tens\extd R) ,\]
\[ \nabla\extd T=-\sqrt{\overline{\delta}}e^{2T\sqrt{\overline{\delta}}}\extd R\tens\extd R\]
modelled on the classical one. We have taken the same Christoffel symbols, just with the quantum tensor product. We then check that this extends as a left quantum connection in the sense $\nabla(f\omega)=\extd f\tens\omega+f\nabla\omega$ for all 1-forms $\omega$ and all elements $f$ of our quantum algebra. Note that in physics we usually think of a connection as a covariant derivative (say, on 1-forms) and along vector fields, but one can think of it equally as a map from  1-forms to a tensor product of 1-forms where the first tensor factor is waiting to evaluate against any vector fields. It is the latter that we take as a definition of $\nabla$ in the quantum case. Torsion-freeness holds in the sense $\wedge\nabla\extd R=\wedge\nabla\extd T=0$ under the wedge product (here $\extd T,\extd R$ anticommute as usual from the fact that the $\extd t,\extd r$ do). 

We also have a right hand connection rule $\nabla(\omega f)=\sigma(\omega\tens\extd f)+(\nabla\omega).f$ as in \cite{DV2,Mou,BegMa6,BegMa5} where, in our case, $\sigma$ is the `flip' map on $\extd R,\extd T$. These values of $\sigma$ are determined from $\nabla$ by the formula stated, we just have to check that it is well-defined when  extended `strongly tensorially' as a bimodule map i.e. commuting with multiplication by elements of the quantum coordinate algebra from either side. On general 1-forms it won't simply be a flip. For example, using the commutation relations for the calculus,
\begin{eqnarray*} \sigma(\extd R R\tens\extd R)&=&\sigma(\extd R\tens R\extd R)\\
&=&\sigma(\extd R\tens\extd R.R)+\lambda'\sqrt{\overline{\delta}}\sigma(\extd R\tens\extd R)\\
&=&\sigma(\extd R\tens\extd R)R+\lambda'\sqrt{\overline{\delta}}\sigma(\extd R\tens\extd R)\\
&=&\extd R\tens\extd R(R+\lambda'\sqrt{\overline{\delta}}).\end{eqnarray*}
We also have $\nabla$ `real' in the sense \cite{BegMa6,BegMa5}
\[ \nabla(\omega^*)=\sigma((*\tens *){\rm flip}\nabla\omega)\]
for all 1-forms $\omega$. Finally,  metric compatibility now makes sense in the form $\nabla g=0$ where $\nabla$ is computed by the rules above and extended to the two tensor factors in $g$ by acting on each factor and using $\sigma$ to flip the left output of $\nabla$ to the far left.  We compute
\begin{eqnarray*}\nabla g&=&\nabla(e^{2T\sqrt{\overline{\delta}}}\extd R)\tens\extd R\\
&&-\sqrt{\overline{\delta}}\sigma(e^{2T\sqrt{\overline{\delta}}}\extd R\tens\extd R)\tens\extd T\\
&&-\sqrt{\overline{\delta}}\sigma(e^{2T\sqrt{\overline{\delta}}}\extd R\tens\extd T)\tens\extd R\\
&&-\nabla\extd T\tens\extd T+\sqrt{\overline{\delta}}\sigma(\extd T\tens e^{2T\sqrt{\overline{\delta}}}\extd R)\tens\extd R\\
&=&0.
\end{eqnarray*}
Here the value of $\nabla$ on the second tensor factor has been inserted and $\sigma$ is used to bring its left output to the far left. When the value of $\nabla$ on the first tensor factor is also inserted and the rules for $\sigma$ are used, we find all the terms cancel and we get zero. The curvature $R_\nabla$ as a 2-form valued operator on 1-forms can also be computed using the definitions in \cite{BegMa5,BegMa6} and one finds
\[ R_\nabla\extd R=\overline{\delta}\extd R\wedge\extd T\tens\extd T,\quad R_\nabla\extd T=\overline{\delta} e^{2T\sqrt{\overline{\delta}}}\extd R\wedge\extd T\tens\extd R.\]
Lifting the 2-forms to antisymmetric tensors and tracing, one then gets ${\rm Ricci}=\overline{\delta} g$ when normalised in a way that matches the classical conventions. For this, the inverse metric is
$\(\extd R,\extd R)=e^{-2T\sqrt{\overline{\delta}}}$, $(\extd T,\extd T)=-1$  extended as a bimodule map in the manner explained in Section~\ref{Sec2}.
These calculations for `quantum de Sitter geometry' would be much harder in the $r,t$ algebra variables but in the $R,T$ ones, which are very close to classical, we see that they follow the classical form provided we are careful about some of the orderings. 

The general case in $n\ge 4$ for the `quantum Bertotti-Robinson space' is not really any different. In the quantum case we do not want to work with angles but work with $z^i={x^i\over r}$. These commute with $r,t$ and, in the $\alpha$ calculus, so do their differentials $\extd z^i=\omega^i/r$ as we saw in Section~\ref{Sec3.1}. Hence they describe an entirely classical $S^{n-2}$ which commutes with  $R,T$ as well. After our change of variables the quantum metric in Section~\ref{Sec3.1} now becomes
\[ g=\delta^{-1}\extd\Omega^2+e^{2T\sqrt{\overline{\delta}}}\extd R\tens\extd R-\extd T\tens\extd T\]
much as before, with the quantum Levi-Civita connection then following similarly to the above. In principle, there may  also be exotic quantum Levi-Civita connections with no classical limit, a phenomenon seen in the model in \cite{BegMa6}. We have only here done the algebraic level and there may be issues at the operator algebras level. At that point the model may usefully tie up with a different approach to noncommutative geometry in \cite{Con}. 
\goodbreak

\end{document}